\documentclass[prl,twocolumn,showpacs,superscriptadress]{revtex4}
\usepackage{amsmath}
\usepackage{amssymb}
\usepackage[dvips]{graphicx}
\usepackage{dcolumn}
\usepackage{color}
\begin{document}
\title{Superconductivity in Semiconductor Structures: the Excitonic Mechanism%
}

\author{E. D. Cherotchenko$^1$}
\author{T. Espinosa-Ortega$^2$}
\author{A. V. Nalitov$^{3}$}
\author{I. A. Shelykh$^{2,4}$}
\author{A. V. Kavokin$^{1,5}$}
\affiliation{$^1$Physics and Astronomy School, University of Southampton, Southampton,
SO171BJ, UK}
\affiliation{$^2$Division of Physics and Applied Physics, Nanyang Technological
University, Singapore 637371.}
\affiliation{$^3$Institut Pascal, PHOTON-N2, Clermont Universit\'{e}, Blaise Pascal
University, CNRS, 24 avenue des Landais, 63177 Aubi\`{e}re Cedex, France.}
\affiliation{$^4$Science Institute, University of Iceland, Dunhagi-3, IS-107 Reykjavik,
Iceland.}
\affiliation{$^5$Spin Optics Laboratory, State University of Saint-Petersburg, 1,
Ulianovskaya, St-Petersburg, Russia.}

\begin{abstract}
We study theoretically the effect of the fermion and boson densities on the superconductivity transition critical temperature $(T_c)$ of a two dimensional electron gas (2DEG), where superconductivity is mediated by a Bose-Einstein condensate of exciton-polaritons. 
The critical temperature is
found to increase with the boson density, but surprisingly it decreases with the 2DEG density increase. 
This makes doped semiconductor structures with shallow Fermi energies better adapted for observation of the exciton-induced superconductivity than metallic layers.
 For the realistic GaAs-based
microcavities containing-doped and neutral quantum wells we estimate $T_c$
as close to 50K.
 Superconductivity is suppressed by magnetic fields of the
order of 4T due to the Fermi surface renormalisation.
\end{abstract}

\maketitle

High temperature superconductivity (HTSC) has been desperately searched for
during decades since the appearance of the seminal work of
Bardeen-Cooper-Schrieffer (BCS)
\cite{Bardeen1957} in the early 50’s. 
Among many different paths physicist have tried to
achieve it, the excitonic mechanism of superconductivity (SC) deserves a
particular attention\cite{Little1964,GinzburgBook,MorelAnderson}.
According to Ginsburg 
\cite{Ginzburg1970,Ginzburg1976}, excitons are expected to be suitable for
realization of HTSC because the characteristic energy above which the
electron attraction mediated by excitons vanishes is several orders of
magnitude larger than the Debye energy limiting the attraction mediated by
phonons.

Despite optimistic expectations, to the best of our knowledge, the exciton
mechanism of SC has never worked until now, most likely due the reduced
retardation effect\cite{MorelAnderson,RetardationEffect}.
 Phonons in the BCS model are very slow compared to
electrons on the Fermi surface. 
Hence there is a strong retardation effect
in phonon-mediated electron-electron attraction, so that the size of a
Cooper pair is very large (of the order of 100nm), and the Coulomb repulsion
can be neglected at such distances.  
In contrast, an exciton is a very fast
quasi-particle once it is accelerated to the wave-vectors comparable with
the Fermi wave-vector in a metal.   
Therefore the replacement of phonons by
excitons leads to the loss of retardation and the smaller sizes of Cooper
pairs, that is why the Coulomb repulsion starts playing an important role.
In realistic multilayer structures the Coulomb repulsion appears to be
stronger than the exciton-mediated attraction so that Cooper pairs cannot be
formed.
In literature \cite{Aruta50KCuprates,Gozar2008} one find reports on
layered metal-insulator structures where SC at 50K in layered
metal-insulator structures, nevertheless there is still no evidence that the
excitonic mechanism is responsible for this effect.

Recently, the novel mechanism to achieve superconductivity mediated by
exciton-polaritons has been proposed in references \cite{FabricePRL,FabriceLongPaper}.
Exciton-polaritons are quasiparticles that
arise due to the strong coupling of excitons with light.
Particularly
interesting exciton-polariton related phenomena have been observed in
semiconductor quantum wells (QW) embedded in microcavity\cite%
{Microcavities,Imamoglu1996}.
Bose-Einstein condensation of cavity
exciton-polaritons at room temperature has been observed \cite%
{Deng2010,Plumhof2014,Christ2007,Baumberg2008}, making the exciton-polariton
a promising boson to bind the Cooper pairs at high temperatures.
Moreover,
it has been proved that the strength of electron-electron interactions
mediated by a condensate of exciton-polaritons can be controlled optically.

The systems considered previously in references \cite%
{FabricePRL,FabriceLongPaper} consist of microcavities where free electrons
in a thin layer interact with contained in adjacent semiconductor layer
exciton-polaritons.
Both layers are brought sufficiently close to each other
to assure efficient coupling between the electrons and exciton-polaritons.
In this way, phonons are replaced by excitations of an exciton-polariton
condensate providing exciton-mediated attraction of free electrons.
While
the retardation effect characteristic of the weak-coupling BCS model is
essentially suppressed also in this regime, the exciton-mediated attraction
appears to be strong enough to overcome the Coulomb repulsion for Cooper
pairs of a characteristic size of 10 nm.
In comparison to the mechanism
considered by Bardeen \cite{Bardeen1957} and Ginzburg \cite%
{Ginzburg1970,Ginzburg1976}, electron-electron attraction mediated by
excitons is much stronger in the presence of the exciton-polariton bosonic
condensate for two reasons: first, the exchange energy needed for creation
of an excited state of the condensate is much smaller than the energy needed
to create a virtual exciton.
Second, the exciton-electron interaction
strength increases proportionally to the occupation number of the
condensate.
This exciton-polariton mechanism of superconductivity was
studied theoretically in a model structure where the electron-electron
attraction potential was calculated and then substituted into the gap
equation that yielded the critical temperature of the superconductivity
phase transition.  
The proof of concept calculation showed a high
potentiality of the excitonic mechanism of SC. 

In order to proceed with the experimental verifications of this prediction,
several issues still need to be clarified. 
Namely, it has been unclear how
the concentration of electrons influences the $T_c$ and what structure is
the most appropriate for experimental observation of the predicted effect:
one where the metallic layer is put in contact with the semiconductor, or an
entirely semiconductor multilayer structure containing doped and undoped
QWs. 
\begin{figure}[h] \label{Structure}
\includegraphics[scale=0.415]{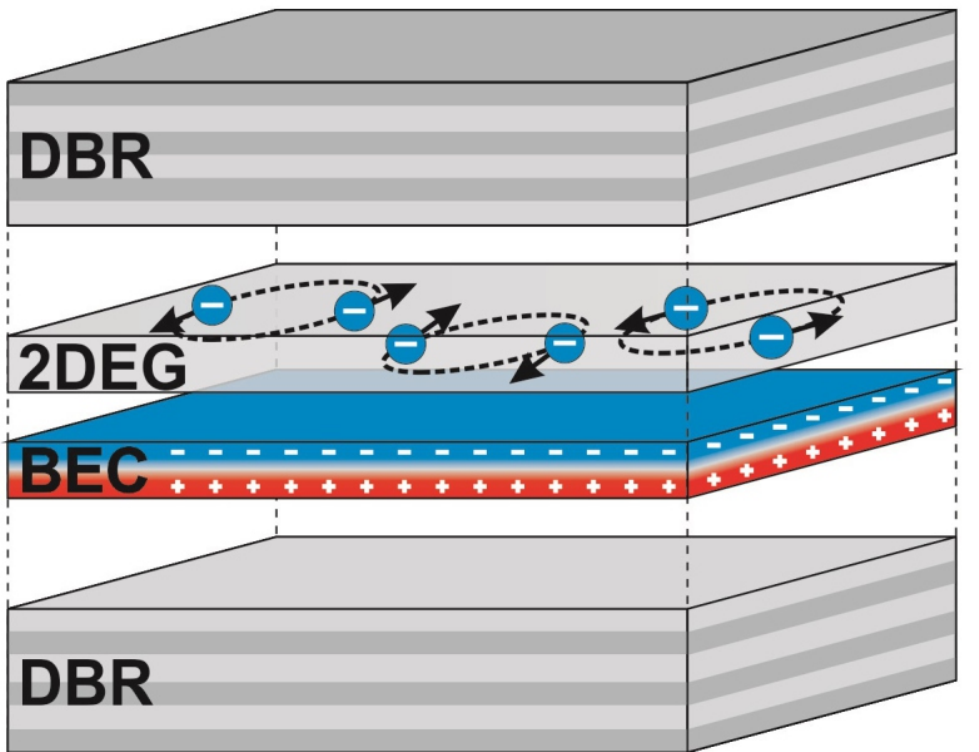} 
\caption{The scheme of the model microcavity structure with an n-dopped QW interacting with an exciton-polariton BEC localized in an adjacent QW.}
\end{figure}

In this Letter, we analyse the behaviour of superconducting gap and $T_c$ as
a function of exciton-polariton and electron concentrations and conclude on
the most convenient structure design for observation of the exciton-mediated
SC. 

The system we study is a microcavity where an electron gas confined to a
quantum well (2DEG) interacts with a polariton condensate localized in an
adjacent semiconductor QW, as shown in Fig.1. 
The microscopic Hamiltonian
that describes this system is derived in ref.\cite{FabriceLongPaper}. Here
we only need the expression for the reduced Hamiltonian that appears after
the Bogoliubov transformation and deacribes the coupling of electrons via
excitations of a polariton condensate, so-called bogolons: 
\begin{eqnarray}
H=\sum_{\mathbf{k}} E_{el}(\mathbf{k})\sigma^\dagger_{\mathbf{k}}\sigma_{%
\mathbf{k}}+ \sum_{\mathbf{k}} E_{bog}(\mathbf{k}) {b}^\dagger_{\mathbf{k}}{b%
}_{\mathbf{k}}+ H_c+  \notag \\
\sum_{\mathbf{k,q}} M({\mathbf{q}})\sigma^\dagger_{\mathbf{k}}\sigma_{%
\mathbf{k+q}}({b}^\dagger_{\mathbf{-q}}+{b}_{\mathbf{q}}).
\end{eqnarray}
Here $E_{el}$ is the free electron energy, the bogolon dispersion is given
by the formula: 
\begin{equation}
\ E_{bog}(\mathbf{k})=\sqrt{\tilde{E}_{pol}(\mathbf{k})(\tilde{E}_{pol}(%
\mathbf{k})+2UN_0A})
\end{equation}
where $\tilde{E}_{pol}={E}_{pol}(\mathbf{k})-{E}_{pol}(\mathbf{0})$. $U$ is
a polariton-polariton interaction potential, $N_0$ is the concentration of
exciton-polaritons in the condensate, $A$ is a normalization area. $H_c$ is
the Coulomb repulsion term, $\mathbf{q}=\mathbf{k_1}-\mathbf{k_2}$, where $\mathbf{k_1}$ and $\mathbf{k_2}$ are the momenta of two interacting
electrons at the Fermi surface, $q=\sqrt{2\mathit{{k_F}(1+\cos{\theta})}}$. 
The renormalized bogolon-electron interaction in (1) is given by $M(\mathbf{q
})$.
It is important that $M(\mathbf{q})\sim \sqrt{N_0}$. The exciton
concentration can be controlled by the external optical pumping, which is
why the strength of Cooper coupling in exciton-mediated superconductors may
be tuned in large limits. 

The effective attraction between two electrons is given by the following
expression: 
\begin{equation}  \label{effectiveattraction}
V_A(\mathbf{q},\omega)=\frac{2M(\mathbf{q})^2E_{bog}(\mathbf{q})}{%
(\hbar\omega)^2-E^2_{bog}},
\end{equation}
 where $\hbar\omega={E}_{pol}(\mathbf{k_1+q})-{E}_{pol}(\mathbf{k_1})$ is an
energy of polariton interchange.
The total effective interaction potential
including Coulomb repulsion is 
\begin{equation}
V_{eff}(\omega)=\frac{A\mathcal{N}}{2\pi}\int_0^{2\pi}[{V_A(\mathbf{q}%
,\omega)+V_C(\mathbf{q})]d\theta, }
\end{equation}
where $\mathcal{N}=m_e/{\pi\hbar^2}$, the Coulomb repulsion is given by $%
V_C(q)=e^2/2\epsilon A(\vert\mathbf{q}\vert+\kappa)$, $\kappa$ is the
screening constant.
Eq.(3) shows that the magnitude of the attraction
potential increases linearly with $N_0$.
This is illustrated by Fig.2(a),
where it is clear that the higher $N_0$ is, the higher the magnitude is and
the larger the attraction region is.
On the contrary in Fig.2(b) one can see that the high concentration of electrons leads to  the decreasing magnitude of the negative part of potential that corresponds to the attraction between electrons.
This effect can be observed in a wide range of polariton concentration values.
The only important limitation to this mechanism of SC is   the Mott transition from the exciton (exciton-polariton) condensate to the electron-hole plasma. 
\begin{figure}[h] \label{FIG2}
\includegraphics[scale=0.215]{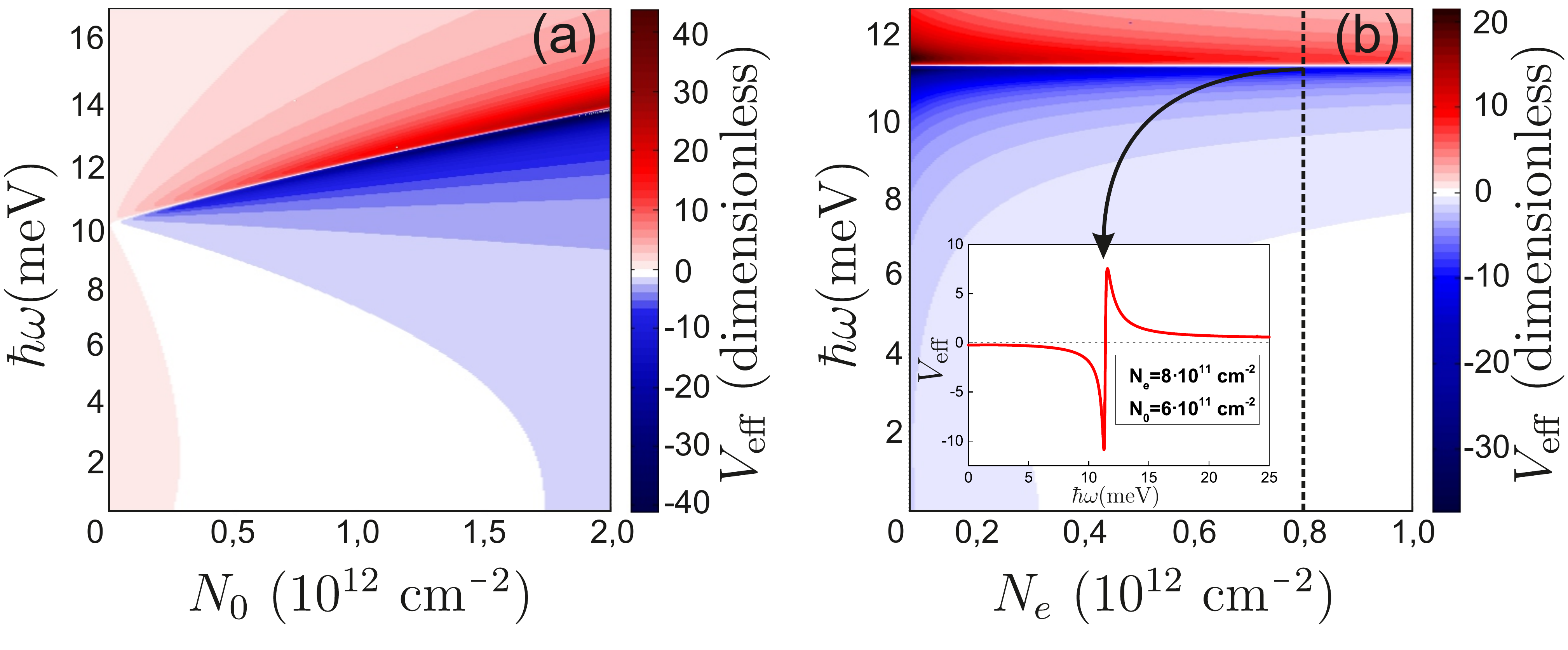} 
\caption{The magnitude of effective interaction potential as a function of a)concentration of polaritons $N_0$ and b)concentration of electrons in 2DEG quantum well.Fig(a)is plotted at the constant $N_e=4\times 10^{11}cm^{-2}$ The color shows the magnitude in dimensionless units.Blue region corresponds to the effective attraction  between electrons, red region represents the repulsion.The inset presents the profile of the potential at the particular concentration $N_e$.} 
\end{figure} 
To obtain the critical temperature of the SC phase transition one needs to
substitute this potential into the gap equation: 
\begin{equation}
\Delta(\omega,\xi)=\int_{-\infty}^{\infty}{\frac{U(\xi-\xi')\Delta(\xi',T)\tanh{(E/2k_BT)}}{2E}d\xi'},
\end{equation}
where $E=\sqrt{\Delta({\xi',T})^2+\xi'^2}$,U is interaction potential.
In the case of a strongly non-monotonous potential $U=V_{eff}(\omega)$  shown in Fig.2 this equation can be solved only numerically. Here we solve it using the iteration method. The example of solution is shown in Fig.3.
\begin{figure}[h] \label{FIG3}
\includegraphics[scale=0.175]{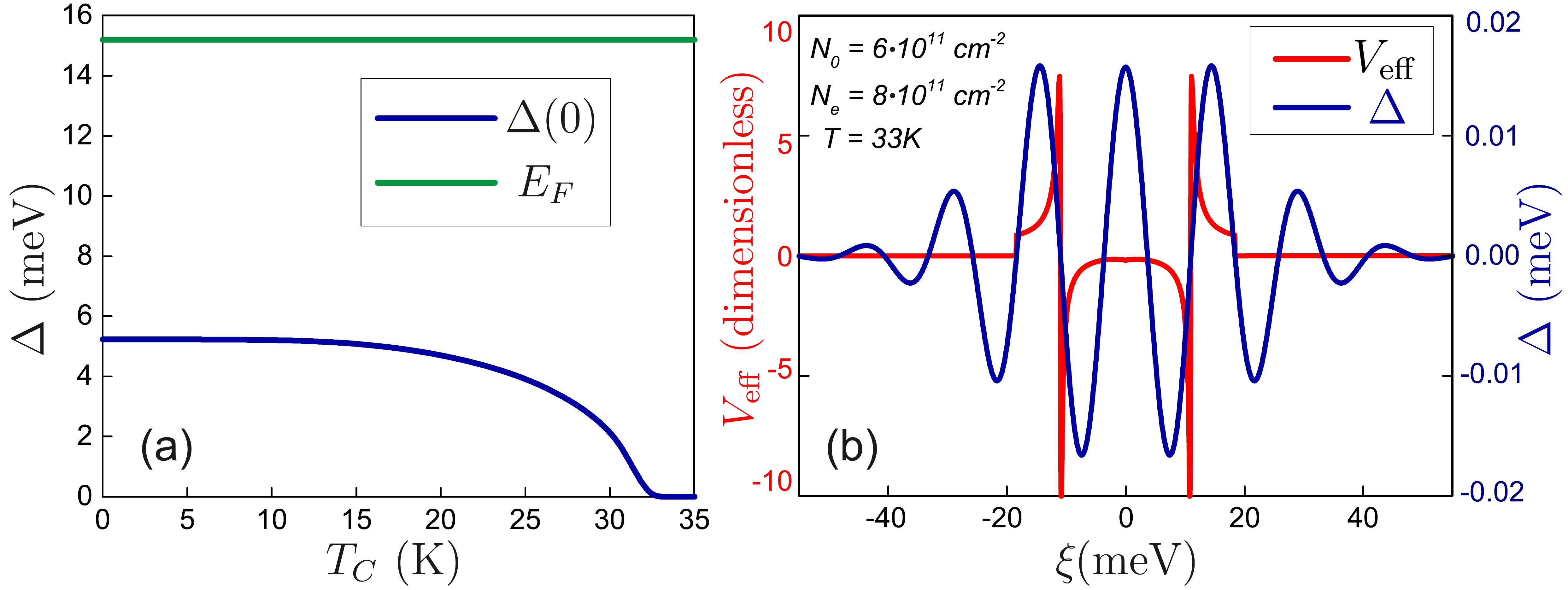} 
\caption{The results of the solution of the gap-equation. Fig.3 (a) shows $\Delta(0)$ as a function of temperature. The critical temperature $T_C$ in this case is equal to 33K. Fig.3 (b) shows  solution of the Eq.(5) at $T=T_C$. The results are presented for the potential with $N_e=8\times 10^{11}cm^{-2}$  and $N_0=6\times 10^{11}cm^{-2}$}
\end{figure} 
We assume that only electrons that are on the Fermi surface can form Cooper pairs. 
Here it means that only the point $\Delta(0)$ has a physical meaning.
If $\Delta(0)>0$, Cooper pair can be formed.
The $T_C$ can be defined as the temperature at which $\Delta(0)$ turns to zero. 

Fig.4 represents the critical temperature of the SC transition as a function of  the concentration of electrons in a 2DEG QW. 
The green line shows the temperature that corresponds to the Fermi energy, the other lines represent the dependences of $T_C$ on the concentration of electrons for the concentration of exciton-polaritons fixed at the different levels.
One can see that increasing $N_e$ leads to the reduction of the critical temperature.
The colored area on the plot shows the range of parameters where our theory is applicable.
\begin{figure}[h] \label{CritTemp}
\includegraphics[scale=0.200]{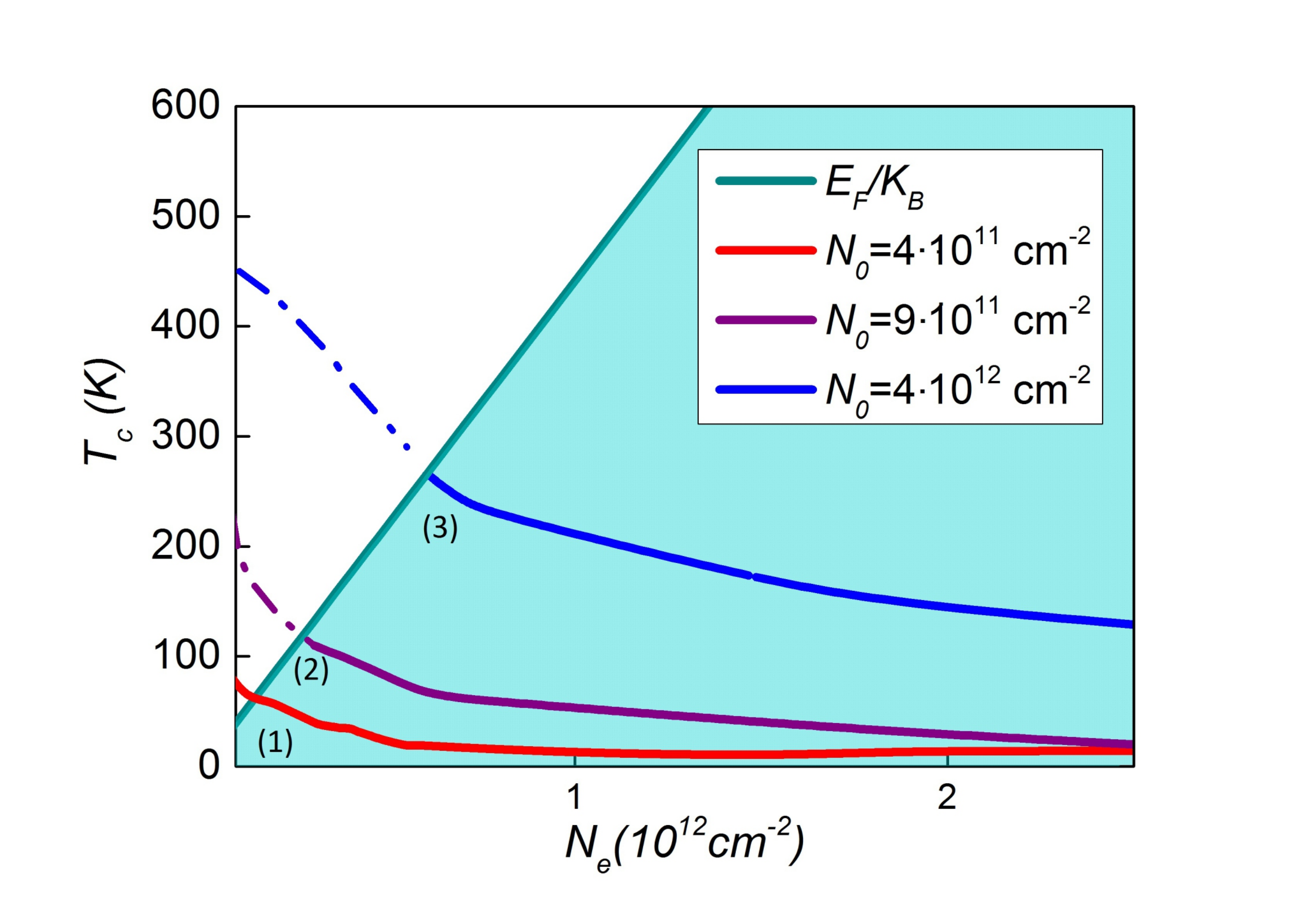} 
\caption{The dependence of $T_C$ on the concentration of electrons in 2DEG QW, plotted for three different polariton concentrations $N_0$. Dashed parts of the curves show the region where the theory is not applicable. Red curves (1,2) represents the parameters of the condensate that are achievable in a realistic GaAs-based semiconductor structures.} 
\end{figure} 

We note that our model has two important limitations.
First,the thermal energy of electrons at the critical temperature must be lower than the Fermi energy.
Otherwise, one cannot assume that electrons obey the Fermi distribution.
Also, the absolute value of the gap-energy must be lower than the Fermi-energy.
In Fig.4  the area of validity of our approach is limited by $E_F=k_BT$ line.
The concentration  $N_0=4\times 10^{12} cm^{-2}$ is apparently beyond the Mott transition threshold, so that it is unrealistic to expect the high $T_C$ predicted by this line.
On the other hand, the exciton concentration  $N_0=4\times 10^{11} cm^{-2}$  is achievable in realistic GaAs based QW structures, so that crictical temperatures of the order of a few tens of Kelvin must be achievable in semiconductor structures.
\bigskip

The superconducting currents may be observed in our structures until the critical current density is achieved. The critical current density
can be conveniently derived from the superconducting gap $\Delta (0)$ as \cite{SSP1996Ibach}

\begin{equation}
j_{c}=\frac{eN_{e}\Delta (0)}{\hbar k_{F}},
\end{equation}

Figure 5(a) shows $j_{c}$ calculated as a function of the
electronic concentration $N_{e}$ and temperature. One can see that the highest current density appears at the lowest concentrations and lowest temperatures on the graph, that fully agrees with our previous calculations.

\bigskip 

Let us discuss now the behaviour of exciton-mediated superconductors in the
presence of external magnetic fields. 
In bulk superconductors, the Meissner
effect exists until the critical magnetic field achieved. This field is linked to the critical current. 
Namely, the critical field induces a surface current equal to $j_{c}$.
At the critical field the superconducting gap is different from
zero.
In our case, the superconducting layer is much thinner than the
typical penetration length of the magnetic field into the superconductor. 
The superconductivity is still
suppressed by the magnetic field in this case, but the gap vanishes at the
critical field $B_{cr}$\cite{DouglassPRL6.346},that can be found from the condition:

\begin{equation}
\Delta (0,B_{cr})=0,
\end{equation}
In order to find $B_{cr}$, we will account for the magnetic field in the gap equation.
The field
affects the radius of the Fermi circle making it larger because of the
decrease of the density of electronic states in a two-dimensional layer. 
A minor effect is the modification of electron-exciton interaction potential
due to the shrinkage of the exciton Bohr radius.
To account for the magnetic field effect on $k_{F}$, we use the expression for the radii of the circles in the reciprocal space, that correspond to Landau levels in the quasiclassical approximation\cite{LandauQM}.
\begin{equation}
k_{p}^2=(p+{1\over 2}){2eB\over\hbar c},  p=0,1,2...
\end{equation}
Electrons may occupy quantum states in the $\Gamma$ vicinity of these circles, where $\Gamma$ is the Dingle broadening of Landau levels dependent on the structural disorder and scattering processes.
The area occupied by electrons in the reciprocal space at each circle at zero temperature can be found as
\begin{equation}
S_{p}=2\pi k_{p}\delta k_{p}
\end{equation}
where $\delta k_{p}={2m\Gamma\over{\hbar^2 k_{p}}}$.
The Fermi wave vector can be expressed as $k_{F}=k_{M}$, where the index $M$ can be found from the condition:
\begin{equation}
{2\over{(2\pi)^2}} {\sum_{p=0}^{M-1}}S_{p}<N_{e}\leq {2\over{(2\pi)^2}}{\sum_{p=0}^{M}}S_{p}
\end{equation}
Fig.5(b) shows $k_{F}$ and $T_{c}$ as functions of magnetic field B for the fixed electron and polariton densities. All parameters are the same that we used for potential calculation for GaAs-structure. In this case $N_e=8\times10^{11}cm^{-2}, N_0=6^\times10^{11}cm^{-2}$,The Dingle broadening of Landau Levels is taken to be to $0.3 meV$. At low magnetic fields given by a condition $\hbar\omega_{c}<\Gamma$ we assume that $k_F=k_F(B=0)$, neglecting the weak oscillations of $k_F$ due to the oscillating electron density of states\cite{ChampelMineev}.
\begin{figure}[h] \label{magfield}
\includegraphics[scale=0.255]{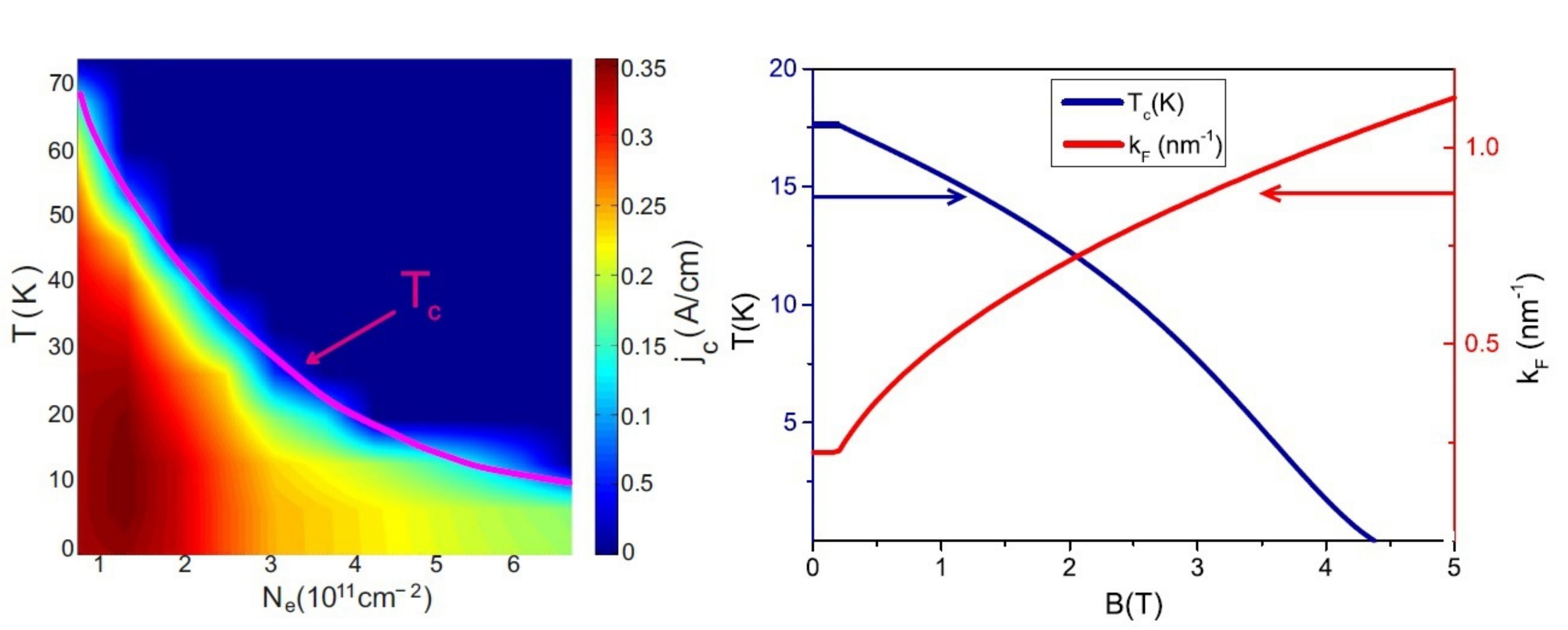} 
\caption{(a) The dependence of the critical current $j_c$ on the temperature and electron concentration. (b)Fermi wavevector (red curve) and critical temperature(blue curve) as a function of magnetic field B.$N_e=8\times10^{11} cm^{-2}$.The Dingle broadening of Landau Levels $\Gamma$ is taken to be 0.3 $meV$, that corresponds to the cyclotron energy $\hbar\omega_c$ at B=0.2T.}
\end{figure}
Our calculation shows that the increase of B at a
constant $\Gamma $ leads to the strong increase of the Fermi wave vector,
which is why the critical temperature decreases and eventually vanishes at $%
B_{c}\approx 4T$.
The increase of $k_{F}$ accounts for the reduction of the
effective area occupied by each electron in the real space due to the
cyclotron motion.
We note, that the validity of the quasi-classical
approximation is limited at strong quantising magnetic fields. In our case,
the number of occupied Landau levels is over 10 even at $B\approx 4T$, which
allows considering the quasi-classical result as a trustworthy
approximation.
Other effects which may influence $B_{c}$ include the
electron Zeeman splitting and edge current effects, which are beyond the
scope of the present work.

In conclusion, multilayer semiconductor heterostructures appear to be promising candidates for the observation of exciton-mediated superconductivity. 
Contrary to the previous expectations, fully semiconductor structures, combining doped and undoped quantum wells provide higher critical temperatures than metal-semiconductor structures.
This can be explained by the fact that
exciton-mediated attraction weakens with the increasing of the Fermi energy faster than the Coulomb repulsion does. 
In the absence of magnetic field we predict the critical temperatures of the order of 50K in realistic GaAs-based microcavities.
We show that external magnetic fields suppress superconductivity in thin semiconductor layers due to the increase of the Fermi wavevector.

\bigskip

Acknowledgements: T.E.-O. and I.A.S. thanks Tier1 project "Polaritons for novel device applications", T.E-O acknowledge Dr. O. Kyriienko
for valuable discussions, E.C. and A.K.thank the EPSRC Established Career Fellowship program for the financial support. A.K. acknowledges support from the Russian Ministery of Science and Education (contract No. 11.G34.31.0067), A.N. acknowledges support from the ITN
INDEX (289968).

\bibliography{reference}

\begin{thebibliography}{21}
\expandafter\ifx\csname natexlab\endcsname\relax\def\natexlab#1{#1}\fi
\expandafter\ifx\csname bibnamefont\endcsname\relax
  \def\bibnamefont#1{#1}\fi
\expandafter\ifx\csname bibfnamefont\endcsname\relax
  \def\bibfnamefont#1{#1}\fi
\expandafter\ifx\csname citenamefont\endcsname\relax
  \def\citenamefont#1{#1}\fi
\expandafter\ifx\csname url\endcsname\relax
  \def\url#1{\texttt{#1}}\fi
\expandafter\ifx\csname urlprefix\endcsname\relax\def\urlprefix{URL }\fi
\providecommand{\bibinfo}[2]{#2}
\providecommand{\eprint}[2][]{\url{#2}}

\bibitem[{\citenamefont{Bardeen et~al.}(1957)\citenamefont{Bardeen, Cooper, and
  Schrieffer}}]{Bardeen1957}
\bibinfo{author}{\bibfnamefont{J.}~\bibnamefont{Bardeen}},
  \bibinfo{author}{\bibfnamefont{L.~N.} \bibnamefont{Cooper}},
  \bibnamefont{and} \bibinfo{author}{\bibfnamefont{J.~R.}
  \bibnamefont{Schrieffer}}, \bibinfo{journal}{Phys. Rev.}
  \textbf{\bibinfo{volume}{106}}, \bibinfo{pages}{162} (\bibinfo{year}{1957}),
  \urlprefix\url{http://link.aps.org/doi/10.1103/PhysRev.106.162}.

\bibitem[{\citenamefont{Little}(1964)}]{Little1964}
\bibinfo{author}{\bibfnamefont{W.~A.} \bibnamefont{Little}},
  \bibinfo{journal}{Phys. Rev.} \textbf{\bibinfo{volume}{134}},
  \bibinfo{pages}{A1416} (\bibinfo{year}{1964}),
  \urlprefix\url{http://link.aps.org/doi/10.1103/PhysRev.134.A1416}.

\bibitem[{\citenamefont{Ginzburg}(2009)}]{GinzburgBook}
\bibinfo{author}{\bibfnamefont{V.~L.} \bibnamefont{Ginzburg}},
  \emph{\bibinfo{title}{On Superconductivity and Superfluidity}}
  (\bibinfo{publisher}{Springer}, \bibinfo{year}{2009}).

\bibitem[{\citenamefont{Morel and Anderson}(1962)}]{MorelAnderson}
\bibinfo{author}{\bibfnamefont{P.}~\bibnamefont{Morel}} \bibnamefont{and}
  \bibinfo{author}{\bibfnamefont{P.~W.} \bibnamefont{Anderson}},
  \bibinfo{journal}{Phys. Rev.} \textbf{\bibinfo{volume}{125}},
  \bibinfo{pages}{1263} (\bibinfo{year}{1962}),
  \urlprefix\url{http://link.aps.org/doi/10.1103/PhysRev.125.1263}.

\bibitem[{\citenamefont{Ginzburg}(1970)}]{Ginzburg1970}
\bibinfo{author}{\bibfnamefont{V.~L.} \bibnamefont{Ginzburg}},
  \bibinfo{journal}{Physics-Uspekhi} \textbf{\bibinfo{volume}{13}},
  \bibinfo{pages}{335} (\bibinfo{year}{1970}),
  \urlprefix\url{http://ufn.ru/en/articles/1970/3/c/}.

\bibitem[{\citenamefont{Ginzburg}(1976)}]{Ginzburg1976}
\bibinfo{author}{\bibfnamefont{V.~L.} \bibnamefont{Ginzburg}},
  \bibinfo{journal}{Physics-Uspekhi} \textbf{\bibinfo{volume}{19}},
  \bibinfo{pages}{174} (\bibinfo{year}{1976}),
  \urlprefix\url{http://ufn.ru/en/articles/1976/2/f/}.

\bibitem[{\citenamefont{Bauer et~al.}(2013)\citenamefont{Bauer, Han, and
  Gunnarsson}}]{RetardationEffect}
\bibinfo{author}{\bibfnamefont{J.}~\bibnamefont{Bauer}},
  \bibinfo{author}{\bibfnamefont{J.~E.} \bibnamefont{Han}}, \bibnamefont{and}
  \bibinfo{author}{\bibfnamefont{O.}~\bibnamefont{Gunnarsson}},
  \bibinfo{journal}{Phys. Rev. B} \textbf{\bibinfo{volume}{87}},
  \bibinfo{pages}{054507} (\bibinfo{year}{2013}),
  \urlprefix\url{http://link.aps.org/doi/10.1103/PhysRevB.87.054507}.

\bibitem[{\citenamefont{Aruta et~al.}(2008)\citenamefont{Aruta, Ghiringhelli,
  Dallera, Fracassi, Medaglia, Tebano, Brookes, Braicovich, and
  Balestrino}}]{Aruta50KCuprates}
\bibinfo{author}{\bibfnamefont{C.}~\bibnamefont{Aruta}},
  \bibinfo{author}{\bibfnamefont{G.}~\bibnamefont{Ghiringhelli}},
  \bibinfo{author}{\bibfnamefont{C.}~\bibnamefont{Dallera}},
  \bibinfo{author}{\bibfnamefont{F.}~\bibnamefont{Fracassi}},
  \bibinfo{author}{\bibfnamefont{P.~G.} \bibnamefont{Medaglia}},
  \bibinfo{author}{\bibfnamefont{A.}~\bibnamefont{Tebano}},
  \bibinfo{author}{\bibfnamefont{N.~B.} \bibnamefont{Brookes}},
  \bibinfo{author}{\bibfnamefont{L.}~\bibnamefont{Braicovich}},
  \bibnamefont{and}
  \bibinfo{author}{\bibfnamefont{G.}~\bibnamefont{Balestrino}},
  \bibinfo{journal}{Phys. Rev. B} \textbf{\bibinfo{volume}{78}},
  \bibinfo{pages}{205120} (\bibinfo{year}{2008}),
  \urlprefix\url{http://link.aps.org/doi/10.1103/PhysRevB.78.205120}.

\bibitem[{\citenamefont{Gozar et~al.}(2008)\citenamefont{Gozar, Logvenov,
  Kourkoutis, Bollinger, Giannuzzi, Muller, and Bozovic}}]{Gozar2008}
\bibinfo{author}{\bibfnamefont{A.}~\bibnamefont{Gozar}},
  \bibinfo{author}{\bibfnamefont{G.}~\bibnamefont{Logvenov}},
  \bibinfo{author}{\bibfnamefont{L.~F.} \bibnamefont{Kourkoutis}},
  \bibinfo{author}{\bibfnamefont{A.~T.} \bibnamefont{Bollinger}},
  \bibinfo{author}{\bibfnamefont{L.~A.} \bibnamefont{Giannuzzi}},
  \bibinfo{author}{\bibfnamefont{D.~A.} \bibnamefont{Muller}},
  \bibnamefont{and} \bibinfo{author}{\bibfnamefont{I.}~\bibnamefont{Bozovic}},
  \bibinfo{journal}{Nature} \textbf{\bibinfo{volume}{455}},
  \bibinfo{pages}{782} (\bibinfo{year}{2008}), ISSN \bibinfo{issn}{0028-0836},
  \urlprefix\url{http://dx.doi.org/10.1038/nature07293}.

\bibitem[{\citenamefont{Laussy et~al.}(2010)\citenamefont{Laussy, Kavokin, and
  Shelykh}}]{FabricePRL}
\bibinfo{author}{\bibfnamefont{F.~P.} \bibnamefont{Laussy}},
  \bibinfo{author}{\bibfnamefont{A.~V.} \bibnamefont{Kavokin}},
  \bibnamefont{and} \bibinfo{author}{\bibfnamefont{I.~A.}
  \bibnamefont{Shelykh}}, \bibinfo{journal}{Phys. Rev. Lett.}
  \textbf{\bibinfo{volume}{104}}, \bibinfo{pages}{106402}
  (\bibinfo{year}{2010}),
  \urlprefix\url{http://link.aps.org/doi/10.1103/PhysRevLett.104.106402}.

\bibitem[{\citenamefont{Laussy et~al.}(2012)\citenamefont{Laussy, Taylor,
  Shelykh, and Kavokin}}]{FabriceLongPaper}
\bibinfo{author}{\bibfnamefont{F.~P.} \bibnamefont{Laussy}},
  \bibinfo{author}{\bibfnamefont{T.}~\bibnamefont{Taylor}},
  \bibinfo{author}{\bibfnamefont{I.~A.} \bibnamefont{Shelykh}},
  \bibnamefont{and} \bibinfo{author}{\bibfnamefont{A.~V.}
  \bibnamefont{Kavokin}}, \bibinfo{journal}{Journal of Nanophotonics}
  \textbf{\bibinfo{volume}{6}}, \bibinfo{pages}{064502} (\bibinfo{year}{2012}),
  \urlprefix\url{http://dx.doi.org/10.1117/1.JNP.6.064502}.

\bibitem[{\citenamefont{Kavokin et~al.}(2007)\citenamefont{Kavokin, Baumberg,
  Malpuech, and Laussy}}]{Microcavities}
\bibinfo{author}{\bibfnamefont{A.~V.} \bibnamefont{Kavokin}},
  \bibinfo{author}{\bibfnamefont{J.~J.} \bibnamefont{Baumberg}},
  \bibinfo{author}{\bibfnamefont{G.}~\bibnamefont{Malpuech}}, \bibnamefont{and}
  \bibinfo{author}{\bibfnamefont{F.~P.} \bibnamefont{Laussy}},
  \emph{\bibinfo{title}{Microcavities}} (\bibinfo{publisher}{Oxford University
  Press, New York}, \bibinfo{year}{2007}).

\bibitem[{\citenamefont{{Imamoglu} et~al.}(1996)\citenamefont{{Imamoglu},
  {Ram}, {Pau}, and {Yamamoto}}}]{Imamoglu1996}
\bibinfo{author}{\bibfnamefont{A.}~\bibnamefont{{Imamoglu}}},
  \bibinfo{author}{\bibfnamefont{R.~J.} \bibnamefont{{Ram}}},
  \bibinfo{author}{\bibfnamefont{S.}~\bibnamefont{{Pau}}}, \bibnamefont{and}
  \bibinfo{author}{\bibfnamefont{Y.}~\bibnamefont{{Yamamoto}}},
  \bibinfo{journal}{Phys. Rev. A} \textbf{\bibinfo{volume}{53}},
  \bibinfo{pages}{4250} (\bibinfo{year}{1996}).

\bibitem[{\citenamefont{Deng et~al.}(2010)\citenamefont{Deng, Haug, and
  Yamamoto}}]{Deng2010}
\bibinfo{author}{\bibfnamefont{H.}~\bibnamefont{Deng}},
  \bibinfo{author}{\bibfnamefont{H.}~\bibnamefont{Haug}}, \bibnamefont{and}
  \bibinfo{author}{\bibfnamefont{Y.}~\bibnamefont{Yamamoto}},
  \bibinfo{journal}{Rev. Mod. Phys.} \textbf{\bibinfo{volume}{82}},
  \bibinfo{pages}{1489} (\bibinfo{year}{2010}),
  \urlprefix\url{http://link.aps.org/doi/10.1103/RevModPhys.82.1489}.

\bibitem[{\citenamefont{Plumhof et~al.}(2014)\citenamefont{Plumhof, Stöferle,
  Mai, Scherf, and Mahrt}}]{Plumhof2014}
\bibinfo{author}{\bibfnamefont{J.~D.} \bibnamefont{Plumhof}},
  \bibinfo{author}{\bibfnamefont{T.}~\bibnamefont{Stöferle}},
  \bibinfo{author}{\bibfnamefont{L.}~\bibnamefont{Mai}},
  \bibinfo{author}{\bibfnamefont{U.}~\bibnamefont{Scherf}}, \bibnamefont{and}
  \bibinfo{author}{\bibfnamefont{R.~F.} \bibnamefont{Mahrt}},
  \bibinfo{journal}{Nat Mater} \textbf{\bibinfo{volume}{13}},
  \bibinfo{pages}{247} (\bibinfo{year}{2014}), ISSN \bibinfo{issn}{1476-1122},
  \urlprefix\url{http://dx.doi.org/10.1038/nmat3825}.

\bibitem[{\citenamefont{Christopoulos et~al.}(2007)\citenamefont{Christopoulos,
  von H\"ogersthal, Grundy, Lagoudakis, Kavokin, Baumberg, Christmann, Butt\'e,
  Feltin, Carlin et~al.}}]{Christ2007}
\bibinfo{author}{\bibfnamefont{S.}~\bibnamefont{Christopoulos}},
  \bibinfo{author}{\bibfnamefont{G.~B.~H.} \bibnamefont{von H\"ogersthal}},
  \bibinfo{author}{\bibfnamefont{A.~J.~D.} \bibnamefont{Grundy}},
  \bibinfo{author}{\bibfnamefont{P.~G.} \bibnamefont{Lagoudakis}},
  \bibinfo{author}{\bibfnamefont{A.~V.} \bibnamefont{Kavokin}},
  \bibinfo{author}{\bibfnamefont{J.~J.} \bibnamefont{Baumberg}},
  \bibinfo{author}{\bibfnamefont{G.}~\bibnamefont{Christmann}},
  \bibinfo{author}{\bibfnamefont{R.}~\bibnamefont{Butt\'e}},
  \bibinfo{author}{\bibfnamefont{E.}~\bibnamefont{Feltin}},
  \bibinfo{author}{\bibfnamefont{J.-F.} \bibnamefont{Carlin}},
  \bibnamefont{et~al.}, \bibinfo{journal}{Phys. Rev. Lett.}
  \textbf{\bibinfo{volume}{98}}, \bibinfo{pages}{126405}
  (\bibinfo{year}{2007}),
  \urlprefix\url{http://link.aps.org/doi/10.1103/PhysRevLett.98.126405}.

\bibitem[{\citenamefont{Baumberg et~al.}(2008)\citenamefont{Baumberg, Kavokin,
  Christopoulos, Grundy, Butt\'e, Christmann, Solnyshkov, Malpuech, Baldassarri
  H\"oger~von H\"ogersthal, Feltin et~al.}}]{Baumberg2008}
\bibinfo{author}{\bibfnamefont{J.~J.} \bibnamefont{Baumberg}},
  \bibinfo{author}{\bibfnamefont{A.~V.} \bibnamefont{Kavokin}},
  \bibinfo{author}{\bibfnamefont{S.}~\bibnamefont{Christopoulos}},
  \bibinfo{author}{\bibfnamefont{A.~J.~D.} \bibnamefont{Grundy}},
  \bibinfo{author}{\bibfnamefont{R.}~\bibnamefont{Butt\'e}},
  \bibinfo{author}{\bibfnamefont{G.}~\bibnamefont{Christmann}},
  \bibinfo{author}{\bibfnamefont{D.~D.} \bibnamefont{Solnyshkov}},
  \bibinfo{author}{\bibfnamefont{G.}~\bibnamefont{Malpuech}},
  \bibinfo{author}{\bibfnamefont{G.}~\bibnamefont{Baldassarri H\"oger~von
  H\"ogersthal}}, \bibinfo{author}{\bibfnamefont{E.}~\bibnamefont{Feltin}},
  \bibnamefont{et~al.}, \bibinfo{journal}{Phys. Rev. Lett.}
  \textbf{\bibinfo{volume}{101}}, \bibinfo{pages}{136409}
  (\bibinfo{year}{2008}),
  \urlprefix\url{http://link.aps.org/doi/10.1103/PhysRevLett.101.136409}.

\bibitem[{\citenamefont{H.Ibach and Luth}(1996)}]{SSP1996Ibach}
\bibinfo{author}{\bibnamefont{H.Ibach}} \bibnamefont{and}
  \bibinfo{author}{\bibfnamefont{H.}~\bibnamefont{Luth}},
  \emph{\bibinfo{title}{Solid State Physics}} (\bibinfo{publisher}{Springer,
  Berlin}, \bibinfo{year}{1996}).

\bibitem[{\citenamefont{Douglass}(1961)}]{DouglassPRL6.346}
\bibinfo{author}{\bibfnamefont{D.~H.} \bibnamefont{Douglass}},
  \bibinfo{journal}{Phys. Rev. Lett.} \textbf{\bibinfo{volume}{6}},
  \bibinfo{pages}{346} (\bibinfo{year}{1961}),
  \urlprefix\url{http://link.aps.org/doi/10.1103/PhysRevLett.6.346}.

\bibitem[{\citenamefont{Landau and Lifshitz}(1977)}]{LandauQM}
\bibinfo{author}{\bibfnamefont{L.~D.} \bibnamefont{Landau}} \bibnamefont{and}
  \bibinfo{author}{\bibfnamefont{E.~L.} \bibnamefont{Lifshitz}},
  \emph{\bibinfo{title}{Quantum Mechanics: Non-Relativistic Theory.}}
  (\bibinfo{publisher}{Pergamon Press.}, \bibinfo{year}{1977}).

\bibitem[{\citenamefont{Champel and Mineev}(2001)}]{ChampelMineev}
\bibinfo{author}{\bibfnamefont{T.}~\bibnamefont{Champel}} \bibnamefont{and}
  \bibinfo{author}{\bibfnamefont{V.~P.} \bibnamefont{Mineev}},
  \bibinfo{journal}{Philosophical Magazine Part B}
  \textbf{\bibinfo{volume}{81}}, \bibinfo{pages}{55} (\bibinfo{year}{2001}).

\end{thebibliography}

\end{document}